\documentclass[prl,aps,twocolumn,floatfix,superscriptaddress,a4paper]{revtex4}

\usepackage{amsmath,amssymb,amsfonts}
\usepackage{bm}
\usepackage{graphicx}

\usepackage{color}

\begin{document}

\title{Edge exponent in the dynamic spin structure factor of the Yang-Gaudin model}

\author{M. B. Zvonarev}
\affiliation{DPMC-MaNEP, University of Geneva, 24 quai Ernest-Ansermet, 1211 Geneva 4, Switzerland}

\author{V. V. Cheianov}
\affiliation{Physics Department, Lancaster University, Lancaster, LA1 4YB, UK}

\author{T. Giamarchi}
\affiliation{DPMC-MaNEP, University of Geneva, 24 quai Ernest-Ansermet, 1211 Geneva 4, Switzerland}

\begin{abstract}
The dynamic spin structure factor $\mathcal{S}(k,\omega)$ of a system of spin-$1/2$ bosons is investigated at arbitrary strength of interparticle repulsion. As a function of $\omega$ it is shown to exhibit a power-law singularity at the
threshold frequency defined by the energy of a magnon at given $k.$ The
power-law exponent is found exactly using a combination of the Bethe Ansatz solution and an effective field theory approach.
\end{abstract}

\date{\today}
\maketitle

The remarkable progress achieved by the theory of
one-dimensional (1D) quantum fluids is rooted in the fact that
dimensionality imposes severe constraints on the fluid's low energy
excitation spectrum. Due to these constraints the investigation of
the low-energy dynamics of the fluid reduces to choosing the effective field theory from a limited number of universality classes. Perhaps the most ubiquitous (and most thoroughly investigated) is
the universality class called the Luttinger
Liquid~\cite{giamarchi_book_1d}. Other non-trivial examples
include states with non-abelian currents,
spin-incoherent \cite{cheianov_spin_decoherent_short,
fiete_spin_decoherent, fiete_SI07} and ferromagnetic
liquids~\cite{zvonarev_ferrobosons07, akhanjee_ferrobosons07,
matveev_isospin_bosons08, kamenev_spinor_bosons08,
zvonarev_BoseHubb08}. For all such cases there exist well
developed analytical methods allowing one to calculate infrared
asymptotics of dynamical correlation function, spectral properties,
and scaling dimensions of local observables.

In a series of recent papers~\cite{zvonarev_ferrobosons07, akhanjee_ferrobosons07, matveev_isospin_bosons08, pustilnik_structfactor_fermions_perturb06, pustilnik_structfactor_calogero06, khodas_FermiLutt07, khodas_structfactor_spinlessbose_lowEfieldtheory07, imambekov_universal08, imambekov_phenomenology08, pereira_structfactor_XXZ_Bethe08, cheianov_structfactor_integrable08, imambekov_structfactor_LiebLiniger_Bethe08, kamenev_spinor_bosons08, zvonarev_BoseHubb08, pereira_structfactor_fermions_Bethe09} it has been found that the dimensionality constraints and the resulting universality may extend far beyond the low energy sector
of the excitation spectrum of the fluid. It was shown that there exist a curve $\omega_-(k)$ in the $(k, \omega)$ space at which spectral functions exhibit power law singularities of the type
\begin{equation}
\mathcal{S}(k,\omega)\simeq c(k)\theta[\omega-\omega_-(k)]
[\omega-\omega_-(k)]^{\Delta(k)}. \label{Athresh}
\end{equation}
Here $\theta(x)$ is the Heaviside step function, $\Delta(k)$ and
$c(k)$ are some momentum-dependent functions, and $\omega_-(k)$ is the energy of the lowest excited state of the fluid at a given momentum $k.$ In
a generic 1D fluid $\omega_-(k)>0$ for all $k$ except a discrete set of points defined by the Luttinger theorem. The spectrum of momentum-dependent anomalous exponents $\Delta(k)$ in Eq.~\eqref{Athresh} is a natural generalization of the spectrum of scaling dimensions of the low-energy effective theory. Understanding the structure of the spectrum of $\Delta(k)$ will greatly
advance the theory of 1D quantum fluids. There are several approaches to this problem. In Refs.~\cite{pustilnik_structfactor_fermions_perturb06, khodas_FermiLutt07} perturbation theory was used to get $\Delta(k)$ in a fermionic system. In Refs.~\cite{khodas_structfactor_spinlessbose_lowEfieldtheory07,
imambekov_universal08, imambekov_phenomenology08} an effective field
theory approach establishing a link with the mobile quantum impurity
problem~\cite{ogawa_mob_impurity92, castella_mob_impurity93, tsukamoto_mob_impurity98} was proposed. This approach was complemented by Bethe Ansatz (BA) calculations for several integrable models:
Calogero-Sutherland~\cite{pustilnik_structfactor_calogero06},
Heisenberg~\cite{pereira_structfactor_XXZ_Bethe08,
cheianov_structfactor_integrable08, pereira_structfactor_fermions_Bethe09}, and Lieb-Liniger~\cite{imambekov_structfactor_LiebLiniger_Bethe08}. Constraints on $\Delta(k)$ implied by symmetries of microscopic Hamiltonian were discussed in Refs.~\cite{kamenev_spinor_bosons08, imambekov_phenomenology08}.

In a recent work~\cite{zvonarev_ferrobosons07} on the dynamical properties of a strongly repulsive ferromagnetic Bose gas observable phenomena such as spin trapping and gaussian damping of spin waves were predicted and a link between these phenomena and the singular behavior, Eq.~\eqref{Athresh}, of the dynamic spin structure factor was established. It was shown that at infinite point-like repulsion and for $k\to 0,$
\begin{equation}
\Delta(k)\simeq -1+ \frac{K}2 \left(\frac{k}{k_F}\right)^2, \label{Deltak0}
\end{equation}
where $K$ is the Luttinger parameter and $k_F=\pi\rho_0$ with $\rho_0$ being average particle density. Assuming the validity of Eq.~\eqref{Deltak0} at a large but finite repulsion, a crossover between trapped and open regimes of spin propagation was characterized completely. A different approach to the dynamics of the same system proposed in Ref.~\cite{akhanjee_ferrobosons07} confirmed Eq.~\eqref{Deltak0}. The approach of Ref.~\cite{akhanjee_ferrobosons07} was further developed in Ref.~\cite{matveev_isospin_bosons08}, demonstrating that for infinite point-like repulsion $\Delta(k)$ has the form~\eqref{Deltak0} for arbitrary $k.$ In Ref.~\cite{kamenev_spinor_bosons08} the small $k$ expansion of $\Delta(k)$ was shown to have the form~\eqref{Deltak0} for arbitrary interparticle repulsion. However, the case of arbitrary $k$ and interparticle repulsion remains unexplored.

In this paper we investigate the behavior of $\Delta(k)$ and $\omega_-(k)$ for the dynamic spin structure factor of a ferromagnetic system of spin-$1/2$ bosons interacting through a point-like repulsive potential of arbitrary strength. This system is described by the Yang-Gaudin model~\cite{gaudin_book}. Combining the Bethe Ansatz with an effective field theory we obtain our main result: explicit expressions \eqref{Deltabeta}--\eqref{Deltaalpha} for $\Delta(k)$ at arbitrary $k$ and interparticle repulsion.

The Hamiltonian of the Yang-Gaudin model is
\begin{equation}
H=
\int_0^L dx\, \left[\partial_x\psi^\dagger_\uparrow\partial_x\psi_\uparrow + \partial_x\psi^\dagger_\downarrow\partial_x\psi_\downarrow + g\rho^2 \right],
\label{Ham1}
\end{equation}
where $\psi_{\uparrow(\downarrow)}(x), \psi_{\uparrow(\downarrow)}^\dagger(x)$ are canonical Bose fields satisfying periodic boundary conditions on a ring of circumference $L,$ and $\rho(x)$ is the total particle density operator. We consider the dynamic spin structure factor
\begin{equation}
\mathcal{S}(k,\omega)= \int dx\,dt\, e^{i(\omega t-kx)} \langle\Uparrow| s_+(x,t)s_-(0,0) |\Uparrow\rangle. \label{specrep}
\end{equation}
Here $s_+(x)=\psi^\dagger_\uparrow(x) \psi_\downarrow(x)$ is the local spin raising operator, and $s_-(x)= [s_+(x)]^\dagger.$ The average in Eq.~\eqref{specrep} is taken with respect to a fully polarized ground state $|\Uparrow\rangle$ of the Hamiltonian~\eqref{Ham1} satisfying $s_+(x)|\Uparrow\rangle=0$ for all $x.$ In the spectral representation Eq.~\eqref{specrep} takes the form
\begin{equation}
\mathcal{S}(k, \omega)=\sum_{f} \delta(\omega-E_f (k)) \vert \langle f, k \vert s_{-}(k)\vert \rm \Uparrow \rangle\vert^2,
\label{specrep2}
\end{equation}
where the sum is taken over the eigenstates $|f,k\rangle$ of the Hamiltonian~\eqref{Ham1} carrying the momentum $k.$ The energies $E_f(k)$ are defined by $H|f,k\rangle= E_f(k)|f,k\rangle.$ The frequency $\omega_-(k)$ in Eq.~\eqref{Athresh} is given by $\omega_-(k)= \min_f E_f(k).$ Thus the calculation of $\omega_-(k)$ reduces to the analysis of the energy spectrum of excitations. The calculation of $\Delta(k)$ directly from the formula~\eqref{specrep2} is a far more difficult task. It requires the knowledge of the matrix element and their resummation procedure. For most integrable models, including Yang-Gaudin, such calculation is beyond the reach of the existing theory. A way to bypass this problem is to combine the BA with an effective field theory. This is the route we take in our calculations.

We begin our analysis with a brief description of BA equations and a calculation of $\omega_-(k).$ All the states $|f,k\rangle$ in Eq.~\eqref{specrep2} lie in the sector with the $z$ projection of the total spin given by $S_z=N/2-1$. In this sector Bethe's wave functions are characterized by
a set of quasimomenta $\{\lambda_1,\ldots,\lambda_N,\xi\}$ which satisfy the BA equations~\cite{gaudin_book}
\begin{equation}
L\lambda_j + \sum_{k=1}^N \theta(\lambda_j-\lambda_k)= 2\pi I_j+\theta(2\lambda_j-2\xi)+\pi. \label{Bethe}
\end{equation}
Here $\theta(\lambda)=2\arctan(\lambda/g)$ is the two-particle phase shift, and $I_j=n_j-(N+1)/2,$ where  $n_j$ are a set of distinct integers. The branch of $\theta(\lambda)$ is chosen so that $\theta(\pm\infty)= \pm\pi.$ The total energy $E$ and momentum $P$ of a system are given by $
E= \sum_{j=1}^N \lambda_j^2$ and $P=\sum_{j=1}^N \lambda_j,$ respectively. The quasimomentum $\xi$ enters in  $E$ and $P$  indirectly, through the solution of Eqs.~\eqref{Bethe}. In the limit $\xi=\infty$ Bethe's equations~\eqref{Bethe} are identical to Bethe's equations of the fully polarized system~\footnote{This fact can be explained by symmetry considerations: The Hamiltonian~\eqref{Ham1} commutes with the total spin lowering operator $S_-.$ By applying $S_-$ to the fully polarized eigenstates of $H$ one gets the eigenstates of the same energy in the sector with $S_z=N/2-1.$}, $S_z=N/2,$
which is equivalent to the Lieb-Liniger model~\cite{lieb_bosons_1D, lieb_excit}. The distribution of $I_j$ in the ground state of the model is
\begin{equation}
I_j= j- \frac{N+1}2, \qquad j=1,\ldots,N. \label{Bethegs}
\end{equation}
Introducing the quasimomenta density $\rho(\lambda_j)=1/[L(\lambda_{j+1}-\lambda_j)]$ and taking the thermodynamic limit $0< \rho_0<\infty$ as $N,L\to\infty$ one gets the integral equation
\begin{equation}
\rho(\lambda)-\frac1{2\pi} \int_{-\Lambda}^\Lambda d\nu\,
\rho(\nu)K(\lambda,\nu)=\frac1{2\pi} \label{Brc1}
\end{equation}
for the quasimomenta in the state~\eqref{Bethegs} and $\xi=\infty$. The kernel $K(\lambda,\nu)\equiv K(\lambda-\nu)$ is $K(\lambda)= \partial\theta(\lambda)/\partial\lambda=2g/(g^2+\lambda^2).$ Note that $\rho(\lambda)$ should satisfy $\int_{-\Lambda}^\Lambda d\lambda\, \rho(\lambda)=\rho_0.$ This formula together with Eq.~\eqref{Brc1} is used to get the value of the Fermi quasimomentum $\Lambda$ as a function of the particle density $\rho_0$. The ground state energy in the thermodynamic limit is
\begin{equation}
E_0= L\int_{-\Lambda}^\Lambda d\lambda\, \lambda^2 \rho(\lambda) \label{E1gs1}
\end{equation}
and the momentum of the ground state is zero.

Consider now the state characterized by a finite value of $\xi$ and $I_j$ given by their ground state values, Eq.~\eqref{Bethegs}. This state is an excitation above the vacuum, which we shall call a magnon. Introducing the so-called shift function~\cite{korepin_book} by $F(\lambda_j|\xi)=(\lambda_j-\tilde\lambda_j)/(\lambda_{j+1}-\lambda_j),$ where $\lambda_j$ are ground state quasimomenta, and $\tilde\lambda_j$ are those of the excited state, we get the following integral equation for $F$ in the thermodynamic limit:
\begin{equation}
F(\lambda|\xi)- \frac1{2\pi}\int_{-\Lambda}^\Lambda d\nu\,
K(\lambda,\nu)F(\nu|\xi)=
-\frac{\pi+\theta(2\lambda-2\xi)}{2\pi}. \label{Fxiint}
\end{equation}
The momentum of the excited state is
\begin{equation}
k= \int_{-\Lambda}^\Lambda
d\lambda\, \rho(\lambda)[\pi+\theta(2\lambda-2\xi)], \label{Dp}
\end{equation}
and its energy above the ground state is
\begin{equation}
\omega_-(k)= -\frac1\pi \int_{-\Lambda}^\Lambda
d\lambda\, \varepsilon(\lambda) K(2\lambda-2\xi).
\label{De}
\end{equation}
Here $\omega_-$ is written as a function of the physical (observable) momentum $k,$ which is related to the quasimomentum $\xi$ by the integral equation~\eqref{Dp}. The quasienergy $\varepsilon(\lambda)$ is given by the solution of the integral equation
\begin{equation}
\varepsilon(\lambda)- \frac1{2\pi} \int_{-\Lambda}^\Lambda d\nu\,
\varepsilon(\nu)K(\lambda,\nu)= \lambda^2-\mu \label{dressed1}
\end{equation}
satisfying a condition $\varepsilon(\pm\Lambda)=0.$
The parameter $\mu$ entering Eq.~\eqref{dressed1} is the chemical potential, defined by $\mu=\left(\partial E_0/ \partial N\right)_L,$ where $E_0$ is found from Eq.~\eqref{E1gs1}. One can show that at small $k$ the dispersion law \eqref{De} is parabolic~\cite{fuchs_spin_waves_1D_bose},
$\omega_-(k)= {k^2}/{2m_*},$ with the effective mass satisfying $
{m_*^{-1} }= -(\pi g\rho_0^2)^{-1}\int_{-\Lambda}^\Lambda d\lambda\, \varepsilon(\lambda) .$

Another way to excite the system is to create a particle-hole pair by moving one of the quantum numbers $I_j$ in \eqref{Bethe} outside of the ground state distribution~\eqref{Bethegs}. Such excitations are analyzed in detail in \cite{lieb_excit} (see also Ref.~\cite{korepin_book}). In particular, at small momentum they are shown to be equivalent to sound waves propagating at velocity
\begin{equation}
v_s= \frac1{2\pi\rho(\Lambda)}\left. \frac{\partial\varepsilon(\lambda)}{\partial\lambda} \right|_{\lambda=\Lambda}. \label{vs}
\end{equation}

\begin{figure}
\includegraphics[width=8 cm]{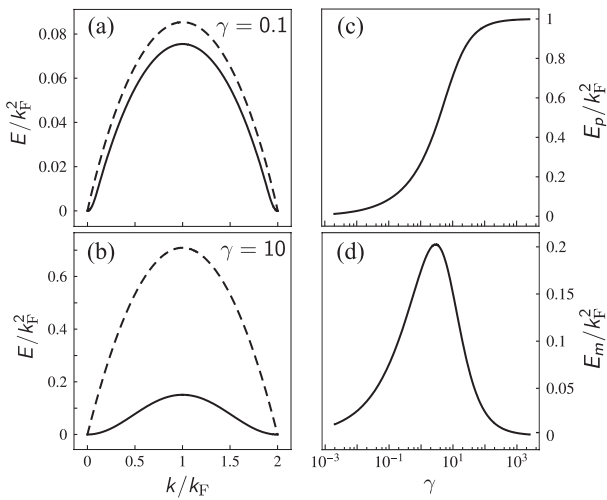}
\caption{Excitation spectrum in the model~\eqref{Ham1}. Panels (a) and (b) show the dispersion curve, $\omega_-(k),$ Eq.~\eqref{De}, of the magnon (solid line) and the exact lower bound of the particle-hole continuum (dashed line) for two values of  $\gamma=g/\rho_0.$ Panel (c) shows the $\gamma$ dependent exact lower bound $E_p$ of the particle-hole continuum at $k=k_F\equiv\pi\rho_0.$ Panel (d) shows the $\gamma$ dependent magnon energy $E_m=\omega_-$ at $k=k_F.$ }
\label{fig:Fig1}
\end{figure}

Any excitation in the $N$-particle sector with $S_z=N/2-1$ consists of several particle-hole pairs and one magnon. The exact lower bound of the particle-hole continuum~\cite{lieb_excit} and the dispersion curve of the magnon are illustrated in Fig.~\ref{fig:Fig1}~(a,b). For all values of coupling constant $g$ the magnon branch lies below the particle-hole continuum.
It is thus single magnon dispersion, Eq.~\eqref{De}, which gives exact lower bound of the excitation spectrum.

While $\omega_-(k)$ is found using BA exclusively, in order to get the threshold exponent, $\Delta(k),$ in Eq.~\eqref{Athresh} for the function~\eqref{specrep} we need to combine the BA solution with a low-energy effective field theory. To do so, we employ the method proposed in \cite{cheianov_structfactor_integrable08}. We introduce an auxiliary microscopic theory with a local Hamiltonian $\tilde H$ depending on $k$ as an external parameter and having the following properties: (i) it conserves the total momentum, which will be denoted by $q.$ (ii) its excitation spectrum at $q=k$ is gapless. (iii) its structure factor $\tilde{\mathcal{S}}$ satisfies
\begin{equation}
\frac{\tilde{\mathcal{S}}(q,\omega)}{\mathcal{S}(q,\omega_-(k)+\omega)} \to 1, \quad q=k, \quad \omega\to 0. \label{AAT}
\end{equation}
In integrable models $\tilde H$ can be constructed as a linear combination of a finite number of mutually commuting local integrals of motion. The eigenstates $|f,q\rangle$ of $H$ are at the same time the eigenstates of $\tilde H,$ therefore $ \tilde{\mathcal{S}}(q, \omega)=\sum_{f} \delta(\omega- \tilde E_f (q)) \vert \langle f, q \vert s_{-}(q)\vert \rm \Uparrow \rangle\vert^2, $
where $\tilde H|f,q\rangle= \tilde E_f(q)|f,q\rangle.$ Like for $H,$ the low energy spectrum of $\tilde H$ consists of sound waves and a magnon. The energy of the magnon is proportional to $(k-q)^2$ as $q\to k.$ The condition~\eqref{AAT} requires that the velocities of the right- and left-moving sound waves be different and given by~\footnote{This condition is analogous to Eq.~(13) of Ref.~\cite{cheianov_structfactor_integrable08}.}
\begin{equation}
v_\pm = v_s \pm \partial\omega_-(k)/\partial k,
\end{equation}
where $v_s$ is given by Eq.~\eqref{vs}.

The dynamics of sound waves is governed by the Luttinger Hamiltonian
\begin{equation}
H_0= \sum_{r=\pm } H_r, \qquad H_r= \frac{v_r}{4\pi} \int_0^L dx :[\partial_x \varphi_r(x)]^2:, \label{Hlutt}
\end{equation}
where the operators $\varphi_r$ are chiral boson fields, $[\varphi_r(x),\varphi_{r^\prime}(x^\prime)]=i\pi r\delta_{rr^\prime}\mathrm{sgn}(x-x^\prime)$
related to the microscopic particle density by
\begin{equation}
\rho(x)=\rho_0+  (2\pi)^{-1}\sqrt K [ \partial_x\varphi_+(x)- \partial_x\varphi_-(x)]
\label{den}
\end{equation}
and the symbol $::$
stands for the boson normal ordering.
In order to describe the low-energy magnon excitation we introduce the spin density field $\tilde {\mathbf s}(x),$  related to the microscopic spin
density by $s_z(x)=\tilde s_z(x) + \rho_0/2$ and $s_\pm (x)=e^{\pm ikx}\tilde s_\pm (x),$ where $s_{\pm}=s_x\pm i s_y$ are the local spin-ladder operators of Eq.~\eqref{specrep}.
Within the effective theory the operators $\tilde s_\pm$ are smooth spin flip fields. Since a local spin flip may excite sound waves, an effective theory should contain a coupling between $\tilde {\mathbf s}$ and $\varphi_\pm.$ The minimal local coupling respecting the $SU(2)$ symmetry of the microscopic theory and vanishing in the absence of magnon excitations~\footnote{The vanishing of the Hamiltonian \eqref{Hi} in the absence of magnon excitations to the leading order in gradient expansion is ensured by the operator identity $\rho(x)=2 s_z(x)$ valid in the fully polarized sector of the system's Hilbert space and by Eq.~\eqref{den}.} is
\begin{equation}
H_i= -\sum_{r=\pm} \frac{v_r\beta_r}{2\pi} \int_0^L dx\, \partial_x \varphi_r(x) \tilde s_z(x).
\label{Hi}
\end{equation}
Other possible couplings involve higher gradient terms, which do not contribute to the critical exponents. The kinetic energy density of the spin field is represented by a higher gradient term  $\partial_x \tilde s_+(x) \partial_x \tilde s_-(x)$ that can also be neglected in the calculation of the critical
exponents \cite{balents_mobimp00}.  The total Hamiltonian of the effective theory describing the dynamics near the threshold is thus given by $H_\text{eff}= H_0 + H_i.$ This Hamiltonian  is diagonalized by a unitary transformation $ e^{ i S} H_\text{eff} e^{-i S}$ with $ S=(2\pi)^{-1} \int_0^L dx [\beta_+\varphi_+(x)-  \beta_-\varphi_-(x)] \tilde s_z(x).$ For the function~\eqref{specrep} this gives
\begin{equation}
\Delta(k)=  - 1 + \frac1{4\pi^2} (\beta_{+}^2+\beta_-^2). \label{Deltabeta}
\end{equation}
What remains is to determine the coupling constants $\beta_{\pm}$ in terms of the parameters of the microscopic theory. This is done by the comparison of the low-energy spectrum of the microscopic Hamiltonian $\tilde H,$ found from the BA solution, and the spectrum of the effective Hamiltonian $H_\text{eff}.$ This procedure yields
\begin{equation}
\beta_r =2\pi r F(r\Lambda|\xi), \qquad r=\pm 1,
\end{equation}
where $F$ is defined by the solution of the integral equation~\eqref{Fxiint}.
We solve this equation and find $\Delta(k)$ numerically for different values of the coupling constant $\gamma = g/\rho_0.$ For easier comparison with Eq.~\eqref{Deltak0} we represent our result in the form
\begin{equation}
\Delta(k)= -1 + \frac{K}2 \left(\frac{k}{k_F}\right)^2 + \frac{(K-1)^2}{K}\alpha(k), \label{Deltaalpha}
\end{equation}
where $k_F=\pi \rho_0$ and $K=k_F/v_s$ is the Luttinger parameter
calculated using Eq.~\eqref{vs}.
\begin{figure}
\includegraphics[width=8 cm]{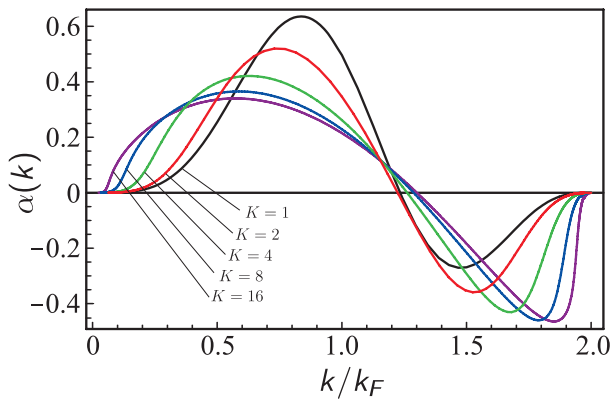}
\caption{The function $\alpha(k)$ defined in Eq. (26) is plotted for different
values of the
dimensionless coupling constant $\gamma.$ The values of the
Luttinger parameter $K$ are indicated for each curve and correspond in
increasing order to $\gamma= \infty,\; 1.65, \; 0.56, \; 0.238 $ and $0.109$
respectively.
}
\label{fig:delta}
\end{figure}
The function $\alpha(k)$ for different values of the Luttinger parameter is shown
in Fig.~\ref{fig:delta}. One can see that at large values of
$\gamma$ (or $K$ close to one) the small $k$ expansion derived in
Ref.~\cite{zvonarev_ferrobosons07} is valid for all values of
$k,$ in agreement with Ref.~\cite{matveev_isospin_bosons08}. It is interesting
to note that the leading correction to this result is of the second order,
$(K-1)^2\sim \gamma^{-2}.$ At arbitrary interaction strength $\alpha(k) \sim k^4$ as $k\to 0,$ therefore the small $k$ expansion of $\Delta(k)$ found in
Ref.~\cite{zvonarev_ferrobosons07} remains valid for all values of
$\gamma,$ confirming the general result of Ref.~\cite{kamenev_spinor_bosons08}. Note that $\alpha(k)$ also vanishes at $k=2k_F.$

The problem considered in the present work is directly related to the X-ray edge problem in the theory of the mobile impurity. In this context, the  model \eqref{Ham1} was investigated in Ref.~\cite{tsukamoto_mob_impurity98}. The approach of Ref.~\cite{tsukamoto_mob_impurity98} exploits a transformation to the co-moving reference frame and combines BA with an effective
field theory similar to ours. The method of Ref.~\cite{tsukamoto_mob_impurity98} has recently been successfully applied
to the Heisenberg model and later was shown \cite{pereira_structfactor_fermions_Bethe09} to produce results equivalent to the method of Ref.~\cite{cheianov_structfactor_integrable08} used here. A direct comparison of the present work with  Ref.~\cite{tsukamoto_mob_impurity98} is however not possible, because the latter used an incorrect BA solution of the model~\eqref{Ham1}.

This work was supported in part by the Swiss National Science Foundation under MaNEP and division II and by ESF under the INSTANS program.

\bibliography{totphys,zvonarev}

\end{document}